\documentclass[sn-mathphys]{sn-jnl}

\usepackage{setspace}
\jyear{2022}%

\theoremstyle{thmstyleone}%
\newtheorem{theorem}{Theorem}
\newtheorem{proposition}[theorem]{Proposition}%

\theoremstyle{thmstyletwo}%
\newtheorem{example}{Example}%
\newtheorem{remark}{Remark}%

\theoremstyle{thmstylethree}%
\newtheorem{definition}{Definition}%

\begin{document}

\title[A long-term formula for a stock price model]{A long-term alternative formula for a stochastic stock price model}


\author*[1]{\fnm{Takuya} \sur{Okabe}}\email{okabe.takuya@shizuoka.ac.jp}

\author*[1,2,3,4,5]{\fnm{Jin} \sur{Yoshimura}}\email{yoshimura.jin@shizuoka.ac.jp}


\affil[1]{\orgdiv{Graduate School of Integrated Science and Technology}, \orgname{Shizuoka University}, \orgaddress{\street{3-5-1 Johoku}, \city{Hamamatsu}, \postcode{432-8561}, \country{Japan}}}

\affil[2]{\orgdiv{Department of International Health and Medical Anthropology, Institute of Tropical Medicine}, 
 \orgname{Nagasaki University}, \orgaddress{\city{Nagasaki}, \postcode{852-8523}, \country{Japan}}}

\affil[3]{\orgdiv{Department of Biological Sciences}, \orgname{Tokyo Metropolitan
University}, \orgaddress{\city{Hachioji}, 
\state{Tokyo}, \postcode{192-0397}, \country{Japan}}}

\affil[4]{\orgdiv{The University Museum}, \orgname{University of Tokyo},
\orgaddress{\city{Bunkyo-ku},\postcode{113-0033}, \state{Tokyo}, \country{Japan}}}

\affil[5]{\orgdiv{Marine Biosystems Research Center}, \orgname{Chiba University},
\orgaddress{\street{Uchiura}, \city{Kamogawa}, \state{Chiba}, \postcode{299-5502}, \country{Japan}}}


\abstract{
This study presents a long-term alternative formula for 
stock price variation described by a geometric Brownian motion 
on the basis of 
median instead of mean or expected values. 
The proposed method is motivated by the observation made in remote fields, where 
optimality of bet-hedging or diversification strategies 
is explained based on a measure different from expected value, like geometric mean. 
When the probability distribution of possible outcomes is 
significantly skewed, 
it is generally known that 
expected value leads to an erroneous picture
owing to its sensitivity to outliers, extreme values of rare occurrence. 
Since geometric mean, or its counterpart median for the log-normal
distribution, does not suffer from this drawback, it provides us
with a more appropriate measure especially for evaluating long-term
outcomes dominated by outliers. 
Thus, the present formula 
makes a more realistic prediction 
for long-term outcomes of a large volatility, for which the probability
distribution becomes conspicuously heavy-tailed. 
}




\maketitle

\subsection*{Article Highlights}
\begin{itemize}

 \item We call into question the validity of expected value concept 
applied to heavy-tailed distributions of stock prices

 \item We propose an alternative formula for evaluating 
future prices based on median

 \item 
The proposed idea is in line with 
biological bet-hedging and the Kelly criterion in optimal betting

%
%
%
%
%
%
%

\end{itemize}

\section{Introduction}

The Black-Scholes-Merton (BSM) theory has been considered the standard
model of prices in financial markets \cite{black73,merton73}.
The Black-Scholes (BS) formula gives the price of a European call
option, i.e., the right to buy a stock on a future day. This formula is
derived under the assumptions of the BSM theory, i.e, a constant
riskless rate, a geometric Brownian motion with constant drift and
volatility, no dividends, no arbitrage opportunity, no commissions and
transactions costs, and a frictionless market \cite{hull18}.
Although the BSM theory has been generally successful and widely
accepted, some shortcomings of the BS formula, including the long-term
prediction, have become clear over the past decades. Many models have
been introduced to overcome the shortcomings originating from the
assumptions of the BSM theory \cite{hull18}. 
Among others, a local volatility model \cite{dupire94} 
and stochastic volatility models \cite{heston93,gatheral06}
are important generalizations to relax the
assumption of constant volatility. 
While these specific points should be addressed on their merits, 
the purpose of the present study is
to approach from a more general perspective to this matter. We cast doubt
on a methodological presupposition in deriving the stock price formula, 
that is, the practicality of the fundamental concept in probability theory, the
expected value. 
In short, we question if the expected value is really expected. 
In various fields facing similar issues, it is generally acknowledged that 
the expected value of random variables 
can happen to deviate significantly from a 
middle of the distribution owing to its strong dependence on outliers,
i.e., extremely large values that occur with extremely small probabilities. 
This strong dependence may cause 
a serious bias toward the overestimation of what is actually expected. 
In particular, expected value 
should deviate appreciably from the expectation of market participants, 
because the future value of stock prices is almost certainly plagued with outliers. 
%
%
Nonetheless, 
expected value is undoubtedly a key concept in economics and finance. 
Thus, we find it worthwhile to seek knowledge from 
remote fields where this problematic situation has been coped with.  

In social and behavioural sciences, it is well known that 
there are cases in which expected value leads to an erroneous result. 
For instance, 
a centuries-old thought experiment of the
St. Petersburg paradox should be mentioned \cite{bernoulli38}.
While the expected value of payoffs is infinite, rational people will not pay more
than a few times as large as the smallest payoff \cite{martin17}.
The point here is just as mentioned above, i.e., 
the expected value has a drawback of 
overestimating the extreme events of rare occurrence.
More specifically, there are abundant examples on this matter 
in the field of ecological biology. 
The number of a population under stochastic environments
varies in a similar fashion as stock prices. 
It is widely acknowledged that the geometric
mean of growth rates (geometric mean fitness) \cite{lewontin69}
provides a more satisfactory picture of population
 growth \cite{yoshimura91,clark00}, 
than the arithmetic mean, i.e., the expected value. 
Risk-spreading or bet-hedging strategies of animals are a concrete example
that is understood by means of geometric mean fitness \cite{yasui18}, 
but not with expected value. 
The geometric mean concept is
intimately related to the median of growth rates, 
as exemplified for the St. Petersburg Paradox \cite{okabe19}. 
The present work is motivated from this observation made in these fields remote from finance.

The structure of the remainder of this paper is as follows. 
In Section 2, we highlight the motivational background by means of 
a multiplicative growth model. This is instructive in two respects. 
Firstly, temporal variation of stock prices is described by a similar model. 
Secondly, this model has already been widely used in ecology of the population growth 
under uncertain environments \cite{lewontin69}, 
as well as in mathematics of optimal strategy in repeated gambles \cite{thorp}. 
The basic idea behind the prior studies is to use 
the geometric mean of randomly varying quantities as a proper measure. 
In Section 3, 
we discuss the stock price model of a geometric Brownian
motion to derive the geometric-mean counterpart, i.e., median, of the stock price formula. 
The main result is given in Eq.~(\ref{medianC}). 
Section 4 provides the numerical results obtained by computer simulation as a concrete
example to support the main point of using median in place of expected value.
Section 5 gives discussions.
Section 6 gives conclusions and future research.

%
%
%
%

\section{Background}

The stochastic behavior of stock price is mathematically modelled as a
geometric Brownian motion (GBM) \cite{bachelier00} 
and it has since long been utilized for a wide application \cite{kanagawa05}.
Most notably, the BSM theory has been considered the standard
model of prices in financial markets \cite{black73,merton73}. 
Before discussing the GBM model, we explain 
the basic idea of the present study based on a simpler model.

In the multiplicative growth model, we suppose that population size
$S_t$ at time $t$ grows geometrically, i.e., it varies in the
multiplicative manner $S_{t+1}=l_t S_t$. 
The growth rates $l_t$ at discrete times $t=0, 1, 2, \cdots$ form 
a sequence of independent, identically distributed random
variables, i.e., each obeys the same probability distribution.
In population ecology, 
Lewontin and Cohen used this model to discuss 
population growth under randomly varying environments \cite{lewontin69}.
Around the same time, essentially the same argument was made in a
different context of repeated bets (gambles),  
i.e., in what is called today the Kelly criterion \cite{thorp}. 
If the rates at different times are independent of each
other, the expected value of $S_t$ is given by $E[S_t]=(\mu_l)^t S_0$, where
$\mu_l=E[l_t]$ is the mean (expected value) of $l_t$. If the mean is greater
than unity ($\mu_l>1$), one would expect that the size $S_t$ blows up as $t$
grows. However, this expectation may be subverted almost certainly. For
example, when $l_t$ comprises two possibilities $l=0.5$ and $l=1.7$ occurring
with equal probability, we have $\mu_l=1.1$ so that the expected size grows
as $E[S_{100}]=1.1^{100} S_0=13781 S_0$. Nevertheless, as a matter of fact, the probability
that the final size $S_{100}$ surpasses the original value $S_0$ is very small
(Prob$[S_{100}>S_0]=0.092$) (Fig. 1). Therefore, contrary to the
expectation, population size is almost certain to vanish ($S_{100}\simeq
0$). This
paradox is caused by skewness, or asymmetry, of the probability
distribution above and below the mean (the expected value). 
In fact, the
cumulative probability above the mean is significantly smaller than that
below (Prob$[S_t>E[S]]=0.022$ and Prob$[S_t<E[S]]=0.978$). 
In this problem, 
%
a more satisfactory solution 
is provided by the geometric mean $G=e^{\mu_{\log l}}$ ($S_t\simeq G^t S_0$), 
rather than by the expected value
$\mu_l=E[S_t]$ ($S_t \simeq \mu_l^t S_0$). 
Note that the geometric mean $\sqrt[n]{l_1 l_2 \cdots l_n}$ of $l_1, l_2, \cdots, l_n$ 
is the arithmetic mean in logscale, i.e., $e^{\frac{1}{n}\sum_i \log l_i}$. 
Hence, $G=e^{\mu_{\log l}}$ is the geometric mean. 
Since $\mu_{\log l}=-0.08126<0$, we obtain $G\simeq 0.92$, and 
the most typical behavior is the exponential decay $S_{100}\simeq 0.92^{100} S_0$,
instead of the exponential blow-up $1.1^{100} S_0$ expected from the expected value. 
In fact, it is straightforward to see that 
$M= G^t S_0$ is the median of $S_t$ (see Appendix \ref{medianlognormal}).
We present this example to underscore the antinomy that 
the epected value $E[S_t]=\mu_l^t S_0$ inflates while 
the median $M= G^t S_0$ deflates (Fig. 1).
In this antinomy, the latter 
provides us with a typical behavior 
of our expectation in the sense that 
the latter (median) is almost certain to occur while the former 
(expected value) is little expected (with a 2.2\% probability).

%
%
%
%
%
%
%
%
%

%

%
%
%

\section{Model and Results}

In the multiplicative model, the geometric
mean of randomly varying growth rates corresponds to the median of their 
probability distribution. 
This is generally true for the random variable obeying the lognormal distribution. 
Thus, 
we can make a parallel argument for stock prices because 
they follow the same distribution, the lognormal distribution. 
%
%
%
%


In the BSM model, the market consists of a
risky asset (stock $S$) and a riskless asset (bond $B$). The former obeys
the stochastic differential equation $dS_t=\mu S_t dt+\sigma S_t dW_t$, where
$dW_t$ is a stochastic variable \cite{hobson04}. 
The latter
varies as $B_t=B_0 e^{rt}$ with $B_0=1$. Since the logarithm of the stock
price $S_t$ discounted by $B_t$ follows a normal distribution, the terminal
stock price $S_T$ follows the log-normal distribution $f_{S_T} (x)$, 
i.e., $\log S_T$ obeys the normal distribution with
mean $\log S_0+(r-\sigma^2/2)T$ and variance $\sigma^2 T$, where $T$ is the time to
maturity. For the strike price $K$, the BS formula for the price of a
European call option is given by 
\begin{eqnarray}
C_{BS} (S_0,K,T)&=& 
{e^{-rT}} {\rm max}( E[S_{T}]-K,0) \nonumber\\
&=& e^{-rT} \int_K^\infty ( S_T-K)f_{S_T} (x)dx\nonumber\\
&=&S_0 \Phi(d_1)-e^{-rT} K \Phi(d_2),
\label{C_BS}
\end{eqnarray}
where 
$d_1=(\log(S_0/K)+(r+\sigma^2/2)T)/(\sigma\sqrt{T})$ and
$d_2=d_1-\sigma\sqrt{T}$ \cite{hull18}. 
In a similar manner, it is straightforward to obtain
the median counterpart. 
Noting that the median of the log-normal distribution $f_{S_T} (x)$ 
is given by the exponentiated mean, $S_0 \exp(rT-\sigma^2 T/2)$, we obtain 
%
\begin{equation}
 C= {\rm max}\left(S_0 \exp(-\sigma^2 T/2)-e^{-rT}K, 0 \right), 
\label{medianC}
\end{equation}
i.e., $C=0$ for $K>S_0 \exp((r-\sigma^2/2)T)$
and $C=S_0 \exp(-\sigma^2 T/2)-e^{-rT}K$ otherwise (Fig.2). 
This is the main result of the present study. 
It should be remarked that 
analytical results are available for the expected value as well as the
median 
because the model leads to the log-normal distribution. 
This is not necessarily the case for sophisticated models to take into account practical factors.

%

In the above, we presented a standard method of deriving the BS formula
in which the behavior of stock prices is modelled with the stochastic
differential equation of a geometric Brownian motion. Another original
method is based on solving a diffusion partial differential equation,
the Black-Scholes equation \cite{hull18}. 
The latter method does not serve
the present purpose, not only because it does not deal with individual
processes but the use of expected value is implicitly taken for
granted while referring to a mathematical result on the diffusion
equation.

Generally speaking, the alternative formula $C$ falls below $C_{BS}$,
occasionally significantly, owing to the insensitivity to the extreme
events of rare occurrence that affect the latter. The two formulae $C_{BS}$
and $C$ give the same result in the shortest term, i.e., $C_{BS}=C$ at
$T=0$. However, they make strikingly different long-term predictions
(Fig. 3a). As the time $T$ increases, $C_{BS}$ increases continuously to the
stock price $S_0$, independently of $K$ \cite{black73,merton73}. 
This is not the case for $C$. Accordingly, the $S_0$-dependence of $C$
is modified drastically (Fig.3b). The difference is because the
probability above the mean of the log-normal distribution drops
exponentially as $\Phi(-\sigma \sqrt{T}/2)$ while the median of the log-normal
distribution is independent of $T$. Thus, the present formula resolves
the discrepancy between expected value and typical value that becomes
appreciable in the long-time behaviour, while it coincides with the
original result in case where their difference does not stand out. 

\section{Simulated data}

To support the present method, 
we show numerical results simulated based on 
geometric Brownian motion (GBM) paths (Fig. 4).  If we denote as
$S_{j,t}$ the stock price at the $t$-th time grid of the $j$-th path, the BS
result is given by the mean of the option value at maturity, 
$e^{-rT} {\rm max} ({E}[S_{j,n}]-K,0)$, 
where ${E}[S_{j,n}]=\frac{1}{m} \sum_{j=1}^m S_{j,n}$.
Note that 
$m$ and $n$ are the number of GBM paths and that of time steps, respectively
($S_{j,1}=S_0$ and $S_{j,n}=S_T$ for the $j$-th path).
In Fig. 4, this result 
is indicated with an arrow labeled with mean. 
For comparison, the result by the proposed method, 
the median of the option value at maturity, is 
indicated with an arrow labeled with median.

For the sake of illustration, 
we present $m=100$ paths of price variation in Fig. 4. 
As shown with dotted lines, 
more than two thirds (67 out of 100) end up with no value ($S_{j,n}<K$). 
%
Thus, the median (no value) is more than twice as likely
to result than any positive value. It is important to remark that
about half of the mean value $C_{BS}=6.2$ (i.e., 2.9) comes just from the 5 largest
outcomes 
with $S_T>80$ (standing out in Fig.4). 
We emphasize that these results are typically expected 
unless 
volatility $\sigma$ is assumed negligibly small
and the time $n$ to the maturity is too short. 
The more we increase the path number $m$, 
the more conspicuously the probability to end up with no value increases. 
Accordingly, the probability to obtain `lucky' results
to make a major contribution to the expected value, becomes negligibly small. 
Under these circumstances, 
the median formula is more appropriate and realistic 
than with the expected value approach. 
%
%
%
%
%

%
%
%
%
%

\section{Discussions}

The BS formula is most often used to calculate the market implied
volatility, which is a forward-looking measure that captures the
market's view of the likelihood of changes in an option price. Most of
the assumptions in its evaluation are embedded in the option pricing
model, while others like the target of the present study originate from
methodological presumptions. The present study is aimed at shedding
light on the assumption of the latter kind. The expected value concept
does not properly take account of the likelihood aspect of unevenly
distributed outcomes. The main purpose of this study is to underline the
importance of this aspect, a deep-rooted problem unquestioned so far, by
way of presenting the new formula. In this sense, the present interest
is more of a scientific nature than a practical one. We aimed at
confronting the inadequacy of the expected value $E(S_t)$ when there is an
exponential discrepancy in the ratio of probabilities above and below
it, namely Prob[$S_t>E(S_t)$] and Prob[$S_t<E(S_t)$]. In practice, notable
differences between theoretical and real values are empirically
corrected by introducing ad-hoc parameters as called 'implied
volatility' and previous studies have obtained satisfactory results with
generalized models while based on the expected value
concept \cite{hull18}. 

In probability theory, the expected value (mean, average) is known to be
away from most typical outcomes when the probability distribution has a
large skew or includes a few extreme outliers. Here the median is much
closer to these typical outcomes than the mean. 
Thus, the alternative median formula gives a better estimate for
the call options when there are outliers or large skews, e.g., the
subprime mortgage crisis in 2007-2008 \cite{hull18}. 
Recently, the
log-normal distribution used for stock prices is questioned and the
probability distribution of stock prices have been estimated from the
past records using, for example, the boot-strap model
instead \cite{hull18}. 
We should note that the assumption of the lognormal distribution
can be replaced to such a practical distribution to gain the better
results to avoid the problem of the abovementioned statistical outliers.
In practice, extreme events of a heavy-tailed distribution may have
substantial impacts in many stochastic dynamical systems, including
economical \cite{buldyrev11}, 
and other social
ones \cite{newman}. 
While it is often
hardly possible to quantify them with mathematical rigor, it is still
very important to have a practical scheme as proposed presently that
takes account of the possible effect of extreme events.

\section{Conclusion}

Motivated by bet-hedging strategies in evolutionary ecology 
and optimal betting by means of the Kelly criterion, 
the present study proposes a new conservative long-term formula for 
the stochastic model of stock prices, 
which is based on the median of the log-normal
distribution of future stock prices. 
The presented formula has an advantage over the conventional one in that 
it is insensitive to extremely large values that can be theoretically
possible but practically almost impossible. 
We focused on a European call option 
to underline the practical feasibility of the proposed approach.
Future research directions include a put option, American options, 
and the delta of an option \cite{hull18}, which are immediate, 
and also other fields to deal with exponential stochastic processes, where 
a long-term measure is required for sustainable development. 

%
%
%
%
%
%
%



\begin{appendices}

\section{Median of the log-normal distribution}\label{medianlognormal}

We consider 
$\log S_t=\log S_0+\sum_{t=0}^{t-1} \log l_t$, 
where $l_t$ at $t=0, 1, 2, \cdots$ form 
a sequence of independent, identically distributed random
variables. 
If we denote mean and variance of $\log l_t$ as 
$\mu_{\log l}$ and $\sigma_{\log l}^2$, respectively, 
the central limit theorem states that the mean 
$\frac{1}{t} \sum_{t=0}^{t-1} \log l_t$ for a large $t$ follows the normal distribution
with mean $\mu_{\log l}$ and variance $\sigma_{\log l}^2/t$. 
In orther words, $S_t$ is distributed according to a log-normal distribution. 
Accordingly, the probability
of $S_t$ being less than $K$ is given in terms of 
the cumulative distribution function 
of the standard normal distribution
(with mean 0 and standard deviation 1),  
$\Phi(y) = \frac{1}{\sqrt{2\pi}}
\int_{-\infty}^y e^{-t^2/2} dt$, 
i.e., 
\[
{\rm Prob} [ S_t<K ]
={\rm Prob} \left[ \frac{1}{t} \sum_{t=0}^{t-1} \log
				    l_t <\frac{1}{t} \log \frac{K}{S_0}
				    \right]
=\Phi\left(
\frac{\frac{1}{t}  \log \frac{K}{S_0} -\mu_{\log l}}{\sigma_{\log l} /\sqrt{t} }
\right).
\]
The median $M$ 
is given by the value of $K$ to make this probability equal to 1/2, 
that is, $M=S_0 (e^{\mu_{\log l} } )^t$. 
%
The quantity in the parentheses ($G=e^{\mu_{\log l}}$) is the
geometrical mean of $l$. 
It should be mentioned that
not only the median but other quartiles may serve the present purpose. 
In order to meet the expectation of market participants, 
the representative measure should always keep 
a moderate value to the probability of getting 
$K$ or more, ${\rm Prob} [ S_t>K ]=1-{\rm Prob} [ S_t<K ]$.
The expected value $K=E$ is defective in that it 
makes this probability depend on $t$ and 
 vanish exponentially as $t$ increases.
A rational person would find little value in anything 
with an exponentially small probability.

\end{appendices}

\bmhead{Compliance with Ethical Standards}

This work was partly supported by grants-in-aid from the Japan Society
for Promotion of Science (no. 21K12047 to TO).
The authors have no relevant financial or non-financial interests to
disclose. 
This article does not contain any studies with human participants or animals performed by any of the authors.

The authors have no conflict of interest to disclose.
No datasets were analysed during the current study.

\newpage

\renewcommand{\thefigure}{{\arabic{figure}}}

\begin{figure}[pb]
\centerline{\includegraphics[width=\textwidth]{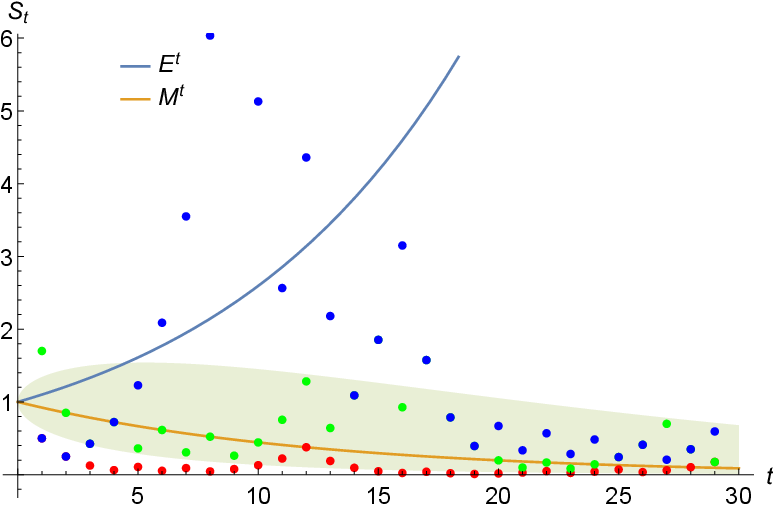}}
\caption{
Temporal variation of size $S_t$ in a stochastic multiplicative growth
 model. The exponential growths by the mean $E$ and the median $M$ are
 contrasted. Dots indicate three realizations of the stochastic
 process. The values within the shaded area account for 68\% of the data
 set (either of the two growth rates $l_1=0.5$ and $l_2=1.7$ occurs with
 a 50\% probability).
}
\end{figure}

\begin{figure}[pb]
\centerline{\includegraphics[width=\textwidth]{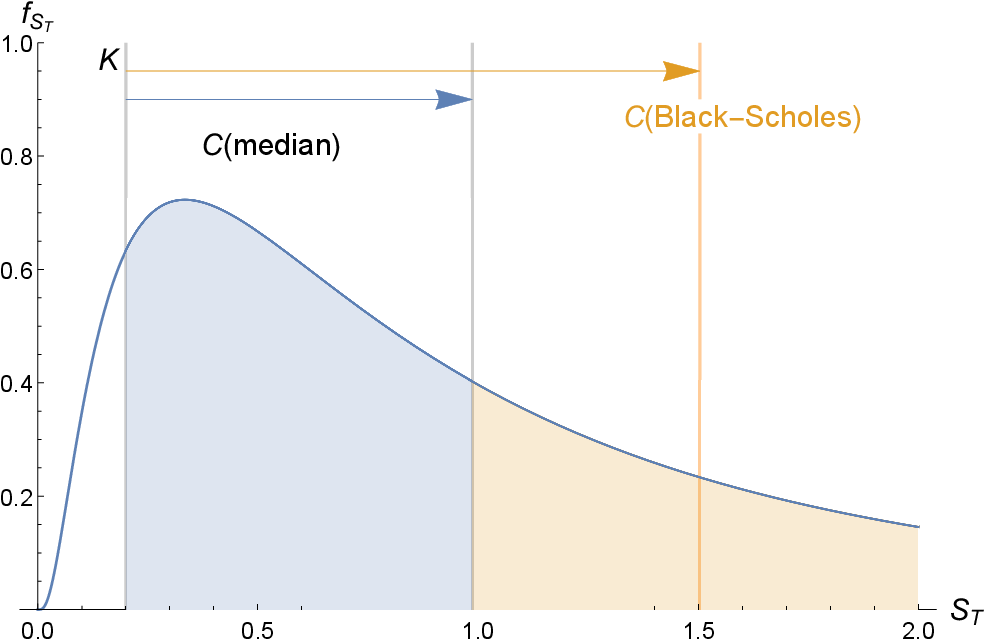}}
\caption{
The median $C$ and the mean $C_{BS}$ (Black-Scholes). The probability
 distribution of price $S_T$ ($S_0=1.5, \sigma=1, T=1, r=0$ and $K=0.2$).
}
\end{figure}

\begin{figure}[pb]
\centerline{\includegraphics[width=\textwidth]{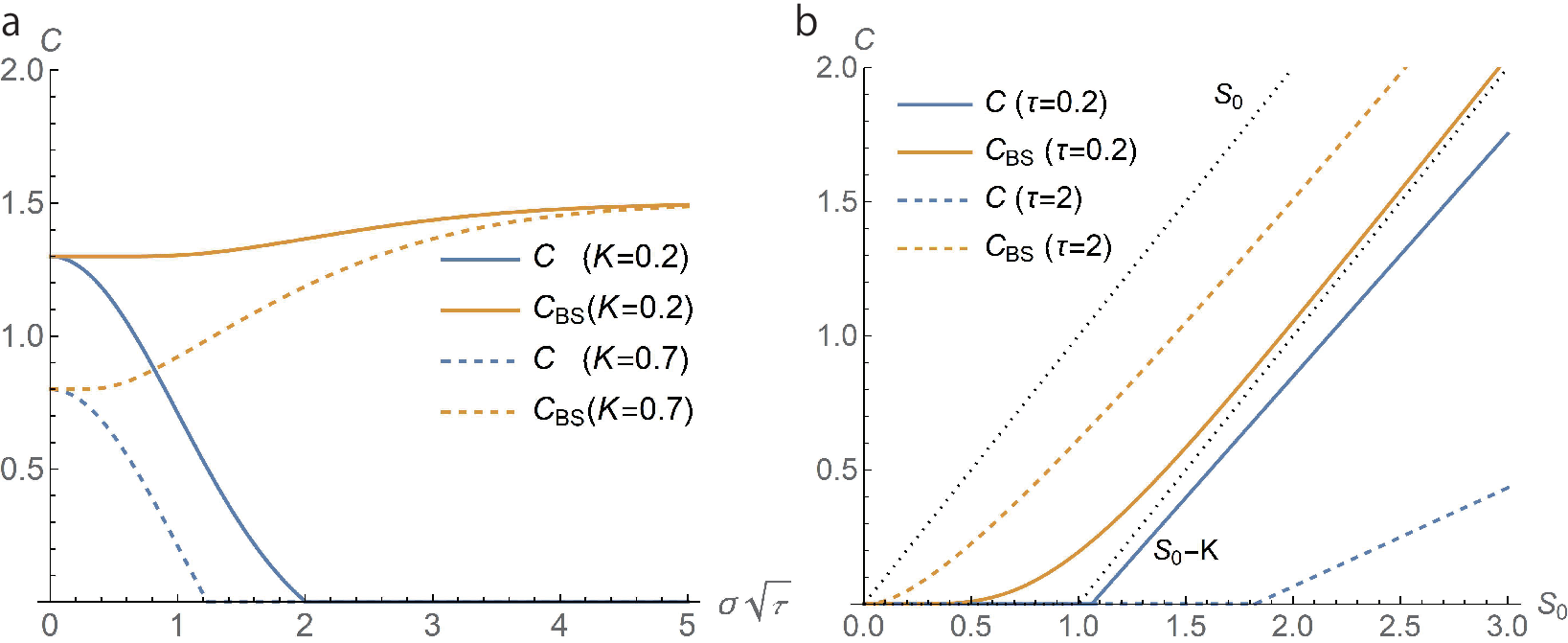}}
\caption{
The median $C$ and the mean $C_{BS}$ (Black-Scholes). a, $C$ and $C_{BS}$ for $K=0.2$
 and 0.7 are plotted against $\sigma\sqrt{T}$ ($S_0=1.5$ and $r=0$). b, $C$ and $C_{BS}$ for
 $T=0.2$ and $T=2$ are plotted against $S_0$ ($r=0.2$, $K=1$ and $\sigma=1$). The BS
 formula $C_{BS}$ gives concave curves between $S_0$ and $S_0-K$.
}
\end{figure}

\begin{figure}[pb]
\centerline{\includegraphics[width=\textwidth]{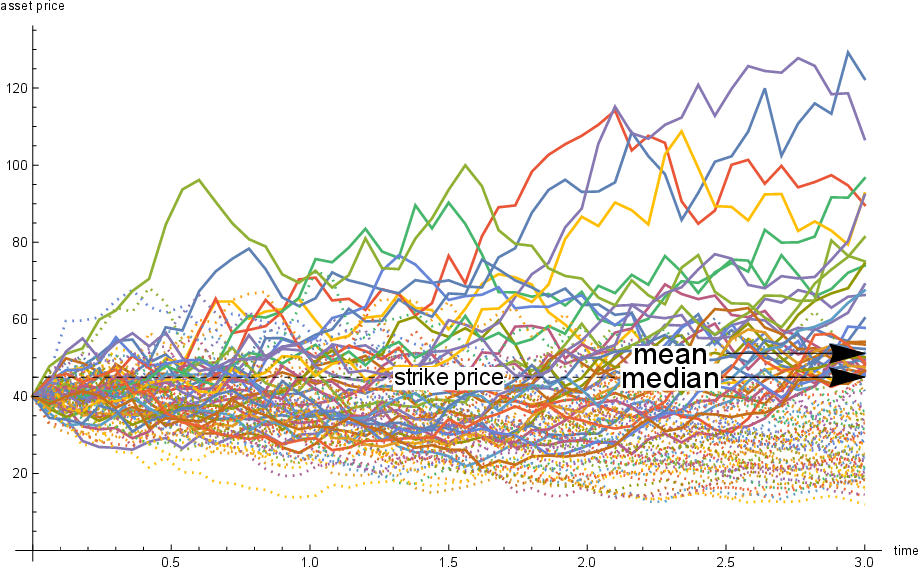}}
\caption{
Temporal variation of stock price $S_t$ as simulated by geometric Brownian
 motion (GBM) paths (starting price $S_0=40$, strike price $K=45$, maturity
 time $T=3$, stock price volatility $\sigma=0.3$, risk-free interest rate
 $r=0.07$, stock dividend yield $\delta=0$, and the number of time steps is
 50). Among the total of 100 simulated paths, 67 paths end up with
 $S_T<K$, which are displayed with dotted lines. An arrow with "mean"
 signifies 
the discounted mean of the option's
 payoffs at expiry. An arrow with "median" is the median counterpart. 
}
\end{figure}

\end{document}